\documentclass[%
 aip,
 amsmath,amssymb,
 reprint,%
]{revtex4-1}

\usepackage{graphicx}% Include figure files
\usepackage{dcolumn}% Align table columns on decimal point
\usepackage{bm}% bold math
\usepackage[utf8]{inputenc}
\usepackage[T1]{fontenc}
\usepackage{mathptmx}
\usepackage{etoolbox}

\makeatletter
\def\@email#1#2{%
 \endgroup
 \patchcmd{\titleblock@produce}
  {\frontmatter@RRAPformat}
  {\frontmatter@RRAPformat{\produce@RRAP{*#1\href{mailto:#2}{#2}}}\frontmatter@RRAPformat}
  {}{}
}%
\makeatother
\begin{document}

\preprint{AIP/123-QED}

\title[]{Non-Hermitian molecular dynamics simulations of %Organic 
exciton-polaritons in lossy cavities}
% Force line breaks with \\
\author{Ilia Sokolovskii}
\affiliation{
  Nanoscience Center and Department of Chemistry, University of
  Jyv\"{a}skyl\"{a}, P.O. Box 35, 40014 Jyv\"{a}skyl\"{a},
  Finland.  }

% \altaffiliation[Also at ]{Physics Department, XYZ University.}%Lines break automatically or can be forced with \\
\author{Gerrit Groenhof}%
 \email{gerrit.x.groenhof@jyu.fi}
\affiliation{
  Nanoscience Center and Department of Chemistry, University of
  Jyv\"{a}skyl\"{a}, P.O. Box 35, 40014 Jyv\"{a}skyl\"{a},
  Finland.  }
%\\This line break forced with \textbackslash\textbackslash
%

\date{\today}

\begin{abstract}
The observation that materials can change their properties when placed inside or near an optical resonator, has sparked a fervid interest in understanding the effects of strong light-matter coupling on molecular dynamics, and several approaches have been proposed to extend the methods of computational chemistry into this regime. Whereas the majority of these approaches have focused on modelling a single molecule coupled to a single cavity mode, changes to chemistry have so far only been  observed experimentally when very many molecules are coupled collectively to multiple modes with short lifetimes. While atomistic simulations of many molecules coupled to multiple cavity modes have been performed with semi-classical molecular dynamics, an explicit description of cavity losses has so far been restricted to simulations in which only a very few molecular degrees of freedom were considered. Here, we have implemented an effective non-Hermitian Hamiltonian to explicitly treat cavity losses in large-scale semi-classical molecular dynamics simulations of organic polaritons and used it to perform both mean-field and surface hopping simulations of polariton relaxation, propagation and energy transfer. 
\end{abstract}

\maketitle

%\begin{quotation}
%The ``lead paragraph'' is encapsulated with the \LaTeX\ 
%\verb+quotation+ environment and is formatted as a single paragraph before the first section heading. 
%(The \verb+quotation+ environment reverts to its usual meaning after the first sectioning command.) 
%Note that numbered references are allowed in the lead paragraph.
%
%The lead paragraph will only be found in an article being prepared for the journal %\textit{Chaos}.
%\end{quotation}

\section{\label{section:intro}Introduction}

%\section{\label{sec:level1}First-level %heading:\protect\\ The line
%break was forced \lowercase{via} %\textbackslash\textbackslash}

Experiments performed in the past decade suggest that placing a material inside an optical micro-cavity or near a plasmonic nano-structure, can change its properties, including energy transfer,\cite{Lerario2017,Zhong2017,Rozenman2018,Myers2018,Zakharko2018,Hou2020,Georgiou2021,Pandya2021,Ostrovskaya2021,Berghuis2022,Balasubrahmaniyam2022,Xu2022,Pandya2022,Huang2023,George2023} charge transport,\cite{Orgiu2015,Krainova2020,Nagarajan2020,Bhatt2021,Fukushima2022} lasing thresholds,\cite{Kena-Cohen2010,Hakala2018bose} and even chemical reactivity.\cite{Hutchison2012,Munkhbat2018, Stranius2018, Thomas2019,Vergauwe2019,Xiong2022} While these changes have been attributed to the hybridization of material and cavity mode excitations into polaritons due to the strong light-matter interaction inside such optical resonators,\cite{Torma2015,Hertzog2019,Garcia-Vidal2021,Rider2022} a lack of theoretical understanding has so far prevented a systematic exploitation of polaritons for controlling the properties of materials. 

% theory approaches based on single molecules 

The established models of quantum optics, such as the Jaynes-Cummings model,\cite{Jaynes1963,Tavis1969} provide conceptual insight into polariton formation, but do not account for the chemical complexity of the molecules nor include the mode structure of the optical resonators, both of which are essential to fully capture and predict how strong light-matter coupling affects the physico-chemical properties of materials. While the theoretical chemistry community has attempted to build strong light-matter coupling into conventional quantum chemistry approaches, \cite{Flick2017a,Haugland2020,Fabri2021,Mandal2023} most of these attempts were aimed at modelling the interaction between a \emph{single} molecule and a \emph{single} confined light mode. In sharp contrast, changes of material properties have so far only been achieved via \emph{collective} strong coupling of very many molecules ({\it{i.e.}}, 10$^5$-10$^8$)\cite{Houdre1996,delPino2015,Eizner2019,Martinez2019} to the quasi-infinite number of confined light modes of an optical resonator.\cite{Hertzog2019,Garcia-Vidal2021} Furthermore, the dramatic changes suggested by calculations for single molecules have not yet been observed experimentally, presumably because the transverse cavity vacuum fields required to bring a single molecule into the strong coupling regime in these calculations, are orders of magnitude higher than the fields inside actual micro-cavities.

% introducing the multiscale MD approach

To go beyond such single-molecule / single-cavity mode description and model the collective interaction between \emph{many} molecules and \emph{multiple} modes of a Fabry-P\'{e}rot micro-cavity in Molecular Dynamics (MD) computer simulations of exciton-polaritons, we had proposed an alternative strategy based on the multiscale quantum mechanics / molecular mechanics (QM/MM) approach.\cite{Luk2017,Tichauer2021} Using extensive parallel computing (millions of atoms on ten thousands CPU cores), we could provide atomistic insights into the dynamics of strongly coupled molecule-cavity systems, including relaxation,\cite{Groenhof2019,Tichauer2021,Tichauer2022} energy transport,\cite{Groenhof2018,Berghuis2022,Sokolovskii2023,Tichauer2023} and photochemistry.\cite{Dutta2023}

% How losses can be included 

In addition to coupling many molecules collectively, optical cavities used in experiments are lossy, and the lifetime of the cavity mode excitations are limited by radiative or Ohmic decay processes. Under the assumption that the cavity modes are weakly coupled to a Markovian bath, such losses can be described with the Lindblad master equation.\cite{Lindblad1976}  Within the single-excitation manifold, which is accessed under the weak driving conditions typically employed in experiments, this master equation can be reformulated in terms of an effective non-Hermitian Hamiltonian,\cite{Visser1995,Saez-Blazquez2017,Echeverri-Arteaga2018} in which the radiative decay of the cavity modes is accounted for by a loss of the norm of the polaritonic wave function. Because in this formalism, the time-evolution of the quantum subsystem is described with the Schr\"{o}dinger equation, it can be directly applied in widely-used semi-classical MD methods, such as Ehrenfest dynamics,\cite{Ehrenfest1927} or Tully's fewest-switches surface hopping.\cite{Tully1990,Tully91,Crespo2018} So far, however, explicit cavity losses, treated either with the Lindblad equation,\cite{Davidsson2020,Torres-Sanchez2021,Koessler2022,Huo2023} or effective non-Hermitian Hamiltonian formalism,\cite{Ulusoy2020,Felicetti2020,Hu2022,Huo2023} have only been included in simulations of molecule-cavity systems with very few molecular degrees of freedom. 

% objective

While in our previous semi-classical simulations of large ensembles of molecules strongly coupled to confined light modes, cavity losses have been modelled \emph{implicitly} as a first-order decay process,\cite{Groenhof2019,Tichauer2021,Sokolovskii2023,Tichauer2023} we have now implemented the effective non-Hermitian Hamiltonian formalism for describing the cavity losses \emph{explicitly} in our multi-scale approach. Because the Hamiltonian of the quantum mechanical subsystem is no longer Hermitian, the adiabatic potential energy surfaces that are normally used in non-adiabatic MD simulations,\cite{Crespo2018} become complex.\cite{Antoniou2020,Kossoski2020,Gyamfi2022} To avoid running classical MD with complex surfaces, we use a hybrid diabatic/adiabatic propagation approach similar to the one proposed by Grannuci {\it{et al.}},\cite{Granucci2001} and already applied within the context of polaritons by Huo and co-workers.\cite{Hu2022,Huo2023} In this hybrid scheme the polaritonic wave function is propagated with the effective non-Hermitian Hamiltonian in the \emph{diabatic representation}, while the classical degrees of freedom representing the nuclei, are evolved under the influence of forces derived from a real \emph{adiabatic potential energy surface}. 

% details about the systems

With the new implementation of the losses we performed both mean-field Ehrenfest and surface hopping MD simulations of molecules strongly coupled to cavity modes with finite lifetimes. To compare between treating losses explicitly via a non-Hermitian Hamiltonian and treating losses implicitly via an {\it ad hoc} first-order decay process, as in our previous works,\cite{Groenhof2019,Tichauer2021,Sokolovskii2023,Tichauer2023} we repeated previous simulations of polariton relaxation and transport in one dimensional (1D) Fabry-P\'{e}rot cavities containing up to 1024 Rhodamine molecules. In addition, we also performed new surface hopping simulations of energy transfer in a hypothetical plasmonic nano-cavity kept together by double-stranded DNA that also contains a Rhodamine dye and a photo-reactive 10-hydroxybenzo[h]quinoline (HBQ) chromophore. Because HBQ can undergo ultra-fast proton transfer into an uncoupled photo-product on timescales comparable to typical cavity mode lifetimes,\cite{Kim2009,Lee2013} there is strong competition between reactive and radiative decay channels in this system. The purpose of the surface hopping simulations therefore is to explore under what excitation conditions the reactive channel dominates.\cite{Perez-Sanchez2023}

The paper is organized as follows: First, in section~\ref{section:theory}, we explain how we include losses explicity into our Tavis-Cummings based multi-scale molecular dynamics model.\cite{Luk2017} Then, in section~\ref{section:simdetails}, we provide the details and parameters of the atomistic simulations of Rhodamine and HBQ coupled to the confined light modes of optical micro- and nano-cavities, followed by a presentation and discussion of the results of these simulations in section~\ref{section:results}. We conclude our paper in section~\ref{section:conclusion} with a  short summary and outlook.

\section{Theoretical Background}\label{section:theory}

\subsection{Effective non-Hermitian Tavis-Cummings Hamiltonian with cavity losses}

To describe the interactions between $N$ molecules and $n_\text{modes}$ lossy cavity modes, we use the Rabi model in the rotating-wave approximation, valid for light-matter coupling strengths below 10\% of the molecular excitation energy,\cite{Forn-Diaz2019} and within the single excitation subspace, valid under the weak driving conditions usually employed in experiments. To account for the radiative decay of the cavity modes, we add the deactivation terms from the Lindblad operator to the Tavis-Cummings Hamiltonian ($\hat{H}^\text{TC}$):
\begin{equation}
\hat{H}= \hat{H}^\text{TC} -\frac{i}{2}\sum^{n_\text{modes}}_k\hbar\gamma_k\hat{a}_k^\dagger\hat{a}_k\label{eq:nonhermham}
\end{equation}
In the second \emph{non-Hermitian} term of this Hamiltonian $\hat{a}_{k}=|0_k\rangle\langle 1_k|$ and $\hat{a}_{k}^\dagger=|1_k\rangle\langle 0_k|$ are the annihilation and creation operators of a photon in cavity mode $k$ with energy $\hslash\omega_k$ and decay rate $\gamma_k$. The first term in Equation~\ref{eq:nonhermham} is the \emph{Hermitian} Tavis-Cummings Hamiltonian,\cite{Jaynes1963,Tavis1969} extended to molecules\cite{Kowalewski2016b,Luk2017} and multiple cavity modes:\cite{Michetti2005,Tichauer2021}
\begin{widetext}
\begin{equation}
\hat{H}^\text{TC}=\sum_j^N\hslash\nu_j({\bf{R}}_j)\hat{\sigma}_j^+\hat{\sigma}_j +\sum_k^{n_\text{modes}}\hbar\omega_k\hat{a}_k^\dagger\hat{a}_k +\sum_k^{n_\text{modes}}\sum_j^N \hbar g_{jk}(\hat{\sigma}_j\hat{a}_k^\dagger + \hat{\sigma}_j^+\hat{a}_k)+\sum_{j}^N V_{\text{S}_0}({\bf{R}}_j)\label{eq:HTC}
\end{equation}
\end{widetext}
Here, $\hat{\sigma}_j^+=|\text{S}^j_1\rangle\langle\text{S}^j_0|$ ($\hat{\sigma}_j=|\text{S}^j_0\rangle\langle\text{S}^j_1|$) is the operator that excites (de-excites) molecule $j$ with excitation energy $h\nu_j(\mathbf R_j)=V_{\text{S}_1}^{\text{mol}}({\bf{R}}_j)-V_{\text{S}_0}^{\text{mol}}({\bf{R}}_j)$ from the electronic ground (excited) state $|\text{S}_0^j({\bf{R}}_j)\rangle$ ($|\text{S}_1^{j}({\bf{R}}_j)\rangle$) to the electronic excited (ground) state $|\text{S}_1^{j}({\bf{R}}_j)\rangle$ ($|\text{S}_0^j({\bf{R}}_j)\rangle$); ${\bf{R}}_j$ is the vector of the Cartesian coordinates of all atoms in molecule $j$; $V_{\text{S}_0}^{\text{mol}}({\bf{R}}_j)$ and $V_{\text{S}_1}^{\text{mol}}({\bf{R}}_j)$ are the adiabatic potential energy surfaces of molecule $j$ in the electronic ground (S$_0$) and excited (S$_1$) state, respectively. The last term in Equation~\ref{eq:HTC} is the total potential energy of the system in the absolute ground state ({\it i.e.}, without any excitation in neither the molecules nor the cavity modes), defined as the sum of the ground-state potential energies of all molecules. As in previous work, we use a hybrid quantum mechanics / molecular mechanics (QM/MM) Hamiltonian~\cite{Warshel1976b, Boggio-Pasqua2012} to model the S$_0$ and S$_1$ potential energy surfaces.\cite{Luk2017} 

In Equation~\ref{eq:HTC} the coupling parameter $g_{jk}$ describes the light-matter interaction between the excitation of molecule $j$ and cavity mode $k$, which within the dipolar approximation depends on the transition dipole moment ${\boldsymbol{\mu}}_j^\text{TDM}$ of the molecule and the vacuum field of the cavity mode:
\begin{equation}
    g_{jk} = -{\boldsymbol{\mu}}_j^\text{TDM}({\bf{R}}_j) \cdot {\bf{f}}_k({\bf{R}}_j) \label{eq:dipole_coupling}
\end{equation}
where vector ${\bf{f}}_k({\bf{R}}_j)$ is the mode function that describes the quantized electromagnetic (EM) field of cavity mode $k$ at the position of molecule $j$. For a Fabry-P\'{e}rot micro-cavity ${\bf{f}}_k({\bf{R}}_j)={\bf{u}}_k\sqrt{\hbar\omega_k/2\epsilon_0 V_\text{cav}}\exp[-i{\bf{k}}\cdot{\bf{R}}_j]$, with ${\bf{u}}_k$ a unit vector indicating the direction of the cavity vacuum field at molecule $j$; {\bf{k}} its two-dimensional $k$-vector; $\epsilon_0$ the vacuum permittivity; and $V_{\text{cav}}$ the mode volume.

\subsubsection{Diabatic basis}

% introduce diabatic basis of product states

Under the assumption that  excitonic interactions between the molecules can be neglected,\cite{Luk2017,Li2023} we compute the elements of the Hamiltonian in Equation~\ref{eq:nonhermham} in the basis of product states between molecular electronic states and cavity mode excitations:
\begin{equation}
\begin{array}{ccl}
|\phi_j\rangle &=& \hat{\sigma}_j^+|\text{S}_0^1\text{S}_0^2..\text{S}_0^{N-1}\text{S}_0^N\rangle\otimes|00..0\rangle\\
\\
&=&\hat{\sigma}_j^+|\Pi_i^N\text{S}_0^i\rangle\otimes|\Pi_k^{n_\text{modes}}0_k\rangle \\
\\
&=&\hat{\sigma}_j^+|\phi_0\rangle
\end{array}\label{eq:basis1}
\end{equation}
for $1\le j\le N$, and 
\begin{equation}
\begin{array}{ccl}
|\phi_{j >N}\rangle &=& \hat{a}_{j-N}^\dagger|\text{S}_0^1\text{S}_0^2..\text{S}_0^{N-1}\text{S}_0^N\rangle\otimes|00..0\rangle\\
\\
&=&\hat{a}_{j-N}^\dagger|\Pi_i^N\text{S}_0^i\rangle\otimes|\Pi_k^{n_\text{modes}}0_k\rangle \\
\\
&=&\hat{a}_{j-N}^\dagger|\phi_0\rangle
\end{array}\label{eq:basis2}
\end{equation}
for $N < j\le N+n_\text{modes}$. In these expressions $|\text{S}_0^i\rangle$ indicates that molecule $i$ is in the electronic ground state, while $|00..0\rangle$ indicates that the Fock states for all $n_\text{modes}$ cavity modes are empty. Thus, the basis state $|\phi_0\rangle$ is the ground state of the molecule-cavity system with no excitation in neither the molecules nor cavity modes:
\begin{equation}
|\phi_0\rangle = |\text{S}_0^1\text{S}_0^2..\text{S}_0^{N-1}\text{S}_0^N\rangle\otimes|00..0\rangle
=|\Pi_i^N\text{S}_0^i\rangle\otimes|\Pi_k^{n_\text{modes}}0_k\rangle\label{eq:phi0}
\end{equation}

Because $|\text{S}_1^j\rangle$ and $|\text{S}_0^j\rangle$ are the orthogonal eigenfunctions of the Hermitian electronic Hamiltonian of bare molecule $j$, the product states in Equations~\ref{eq:basis1} and  \ref{eq:basis2} are also orthogonal. Therefore, the non-adiabatic coupling vector that can drive population transfer between product states with different molecules in the electronic excited state, are zero:
\begin{equation}
{\bf{D}}_{mn}=\langle\phi_m|\nabla_{a\in m}|\phi_n\rangle = \langle\text{S}_0^n|\langle\text{S}_1^m|\nabla_{a\in m} |\text{S}_0^m\rangle|\text{S}_1^n\rangle = 0
\end{equation}
for $m,n \le N$. Here, the gradient is evaluated with respect to a displacement of any atom $a$ in molecule $m$. Furthermore, because the cavity mode Fock states are orthogonal as well, the non-adiabatic coupling vectors ${\bf{D}}_{mn}$ for population transfer between states with and without cavity mode excitation are also zero:
 \begin{equation}
{\bf{D}}_{mn}=\langle\phi_m|\nabla_{a\in m}|\phi_n\rangle = \langle 0_{n-N}|\langle\text{S}_1^m|\nabla_{a\in m}|\text{S}_0^m\rangle |1_{n-N}\rangle = 0
\end{equation}
for $m\le N$ and $n>N$. Thus, within the single-excitation subspace, the product states form a strictly \emph{diabatic} basis. If, in addition to the dynamics in the single-excitation manifold, also the dynamics in the ground state are relevant,\cite{Torres-Sanchez2021} $|\phi_0\rangle$ (Equation~\ref{eq:phi0}) can be included, but in that case, we also need to add the non-adiabatic coupling vectors for internal conversion of the molecules ({\it i.e.}, ${\bf{D}}_{0m}=\langle \text{S}_0^m|\nabla\hat{H}| \text{S}_1^m\rangle$).
 
% dynamics in adiabatic basis vs diabatic basis

\subsection{Ehrenfest dynamics}

% propagator of the wave function

In mean-field, or Ehrenfest, MD the classical degrees of freedom, usually the nuclei, evolve under the influence of the expectation value of forces with respect to the wave function of the quantum degrees of freedom,\cite{Ehrenfest1927} while the wave function evolves along with the classical degrees of freedom. By expanding the total wave function as a linear combination of the \emph{time-independent} diabatic light-matter states (Equations~\ref{eq:basis1} and~\ref{eq:basis2}):\cite{Zhou2022}
\begin{equation}
|\Psi(t)\rangle = \sum_j^{N+n_\text{modes}}|\phi_j\rangle d_j(t)\label{eq:Psi_dia}
\end{equation}
the evolution of the \emph{time-dependent} diabatic expansion coefficients, $d_j(t)$, is obtained by numerically integrating the Schr\"{o}dinger equation over discrete time intervals, $\Delta t$:
\begin{equation}
{\bf{d}}(t+\Delta t)={\bf{P}}^\text{dia}{\bf{d}}(t)\label{eq:SE_dia}
\end{equation}
Here, ${\bf{d}}(t)$ is a vector containing the diabatic expansion coefficients $d_j(t)$ and ${\bf{P}}^\text{dia}$ the propagator in the diabatic basis
\begin{equation}
    {\bf{P}}^\text{dia} =  \exp\left[-i\left({\bf{H}}^\text{TC}(t+\Delta t)+{\bf{H}}^\text{TC}(t)-i\hbar {\boldsymbol{\gamma}}\right)\Delta t/2\hbar\right]\label{eq:propdia}
\end{equation}
with ${\boldsymbol{\gamma}}$ a diagonal matrix containing the cavity mode decay rates, $\gamma_k$, as elements. Because of these decay terms, the norm of the total wave function, $|\Psi(t)|^2=\sum_j^{N+n_\text{modes}}|d_j|^2$, is not conserved but decreases due to the losses.

% gradient expression in diabatic representation

The forces acting on the classical nuclei are computed as  expectation values with respect to the wave function (Equation~\ref{eq:Psi_dia}):
\begin{equation}
\begin{array}{ccl}
    {\bf{F}}_{a}&=&-\nabla_{a}\langle\Psi(t)|\hat{H}|\Psi(t)\rangle/\langle\Psi(t)|\Psi(t)\rangle\\
\\
&=&-\sum_j\sum_k d_j^*(t)d_k(t)\nabla_a\langle\phi_j|\hat{H}|\phi_k\rangle/\sum_i|d_i(t)|^2\label{eq:gradient}
    \end{array}
\end{equation}
for a nucleus $a$ in any of the molecules. For $j\le N$, the diagonal gradient terms inside the double sum are evaluated as:
\begin{equation}
\nabla_{a\in j}\langle\phi_j|\hat{H}|\phi_j\rangle = \nabla_{a\in j} V^\text{mol}_{\text{S}_1}({\bf{R}}_j)
\end{equation}
if atom $a$ belongs to molecule $j$ and
\begin{equation}
\nabla_{a\in i}\langle\phi_j|\hat{H}|\phi_j\rangle = \nabla_{a\in i} V^\text{mol}_{\text{S}_0}({\bf{R}}_{i})
\end{equation}
if atom $a$ does not belong to molecule $j$, but to molecule $i$ instead. For $j > N$, the terms are 
\begin{equation}
\nabla_{a\in i}\langle\phi_j|\hat{H}|\phi_j\rangle = \nabla_{a\in i} V^\text{mol}_{\text{S}_0}({\bf{R}}_i)
\end{equation}
for atom $a$ in any molecule $i$. For atom $a$ in molecule $j$, the gradient of the off-diagonal light-matter coupling term is:
\begin{equation}
\nabla_{a\in j}\langle\phi_j|\hat{H}|\phi_k\rangle =-\nabla_{a\in j}{\boldsymbol{\mu}}^\text{TDM}_j({\bf{R}}_j) \cdot {\bf{f}}_k({\bf{R}}_j)\label{eq:grad_dipole}
\end{equation}
if $j\le N$ and $k > N$. Otherwise, this term is zero. With these terms, the evaluation of the force in Equation~\ref{eq:gradient} simplifies to 
\begin{widetext}
\begin{equation}
    \begin{array}{ccl}
    {\bf{F}}_{a\in j}&=&-\left(|d_j(t)|^2\nabla_{a\in j}V_{\text{S}_1}^\text{mol}({\bf{R}}_j)+\left(n(t)-|d_j(t)|^2\right)\nabla_{a\in j}V_{\text{S}_0}^\text{mol}({\bf{R}}_j)\right)/n(t)\\
    \\
    & & +\left(\sum_{k}^{n_\text{modes}}d_j^*(t)d_k(t)\nabla_{a\in j}{\boldsymbol{\mu}}_j^\text{TDM}({\bf{R}}_j)\cdot{\bf{f}}_k({\bf{R}}_j)\right])/n(t) \\
    \\
    & &+\left(\sum_{k}^{n_\text{modes}}d_k^*(t)d_j(t)\nabla_{a\in j}{\boldsymbol{\mu}}_j^\text{TDM}({\bf{R}}_j)\cdot{\bf{f}}^*_k({\bf{R}}_j)\right)/n(t)\\
    \\
    &=&-\left(|d_j(t)|^2\nabla_{a\in j}V_{\text{S}_1}^\text{mol}({\bf{R}}_j)+\left(n(t)-|d_j(t)|^2\right)\nabla_{a\in j}V_{\text{S}_0}^\text{mol}({\bf{R}}_j)-2\Re\left[\sum_{k}^{n_\text{modes}}d_j^*(t)d_k(t)\nabla_{a\in j}{\boldsymbol{\mu}}_j^\text{TDM}({\bf{R}}_j)\cdot{\bf{f}}_k({\bf{R}}_j)\right]\right)/n(t)
\end{array}\label{eq:linear_scaling_gradient}
\end{equation}
\end{widetext}
where  $n(t)=|\Psi(t)|^2=\sum_i^{N+n_\text{modes}}|d_i(t)|^2$ is the norm of the wave function $\Psi(t)$ at time $t$, and we have used that for complex numbers $z+z^*$$= 2\Re[z]$.

% outline of the Ehrenfest algorithm

Thus, in Ehrenfest MD the matrix representation of the effective non-Hermitian Hamiltonian ${\hat{H}}$ in Equation~\ref{eq:nonhermham} is constructed at every time step of the simulation, using the diabatic basis functions $|\phi_j\rangle$ (Equations~\ref{eq:basis1} and~\ref{eq:basis2}), obtained from QM/MM calculations of the electronic states of the molecules.\cite{Boggio-Pasqua2012} Then, the gradients (Equation~\ref{eq:linear_scaling_gradient}) are computed from the molecular S$_0$, S$_1$ and ${\boldsymbol{\mu}}^\text{TDM}$ gradients in combination with the expansion coefficients ${\bf{d}}(t)$, and used to integrate the positions of the classical nuclei over a time interval, $\Delta t$.\cite{Verlet1967} At the updated nuclear configuration, a new Hamiltonian matrix is constructed  and, after combination with the previous Hamiltonian matrix, used to evolve the expansion coefficients $d_j$ from $t$ to $t+\Delta t$ with the propagator ${\bf{P}}^\text{dia}$ in Equation~\ref{eq:propdia}.

\subsection{Fewest-Switches Surface Hopping }

Because evolution on the mean-field potential energy surface may not always provide a optimal description of the chemical dynamics,\cite{Parandekar2006} so-called "surface hopping" methods have been developed,\cite{Tully1990,Tully91} in which the coherent evolution of the wave function, expanded in a given basis, is combined with the evolution of the classical degrees of freedom on a \emph{single} potential energy surface that is associated with one of the basis states.\cite{Crespo2018} Population transfer between the basis states is modelled by stochastic hops of the classical subsystem between the potential energy surfaces of these states. 

Surface hopping simulations are normally performed in the \emph{adiabatic representation}, in which the basis functions are the eigenstates of the Hamiltonian of the quantum subsystem. However, when the cavity losses are included explicitly, the Hamiltonian is no longer Hermitian, and thus the potential energy surfaces acquire a complex component that is associated with the finite lifetime of the eigenstates.\cite{Antoniou2020} In addition, the eigenvectors are not orthogonal, and although the left and right eigenvectors can be bi-orhtogonalized,\cite{Rosas-Ortiz2018} this nevertheless complicates the evaluation of expectation values. To avoid such issues when running semi-classical MD trajectories on complex potential energy surfaces, we adopt the hybrid diabatic / adiabatic scheme, proposed by Granucci and co-workers,\cite{Granucci2001} in which the wave function is propagated in the \emph{diabatic representation} (Equation~\ref{eq:propdia}), under the influence of the effective non-Hermitian Hamiltonian ({\it{i.e.}}, $\hat{H}$ in Equation~\ref{eq:nonhermham}), while the classical degrees of freedom %dynamics 
evolve on a real adiabatic potential energy surface associated with the eigenstates of the Hermitian part of the total Hamiltonian ({\it{i.e.}}, $\hat{H}^\text{TC}$, Equation~\ref{eq:HTC}).\cite{Hu2022}

The eigenfunctions of $\hat{H}^\text{TC}$ are linear combinations of the diabatic states (Equations~\ref{eq:basis1} and~\ref{eq:basis2}):
\begin{equation}
|\psi^m\rangle = \left(\sum_j^N\beta_j^m\hat{\sigma}_j^++\sum_k^{n_\text{modes}}\alpha_k^m\hat{a}^\dagger_k\right)|\phi_0\rangle=\sum_i^{N+n_\text{modes}}U_{im}|\phi_i\rangle\label{eq:adiabaticbasis}
\end{equation}
with eigenenergies $E_m$. The $\beta_j^m$ and $\alpha_k^m$ are expansion coefficients that reflect the contribution of the molecular excitons ($\vert\text{S}_1^j(\mathbf{R}_j)\rangle$) and the cavity mode excitations ($\vert 1_k\rangle$) to polaritonic eigenstate $\vert \psi^m\rangle$. These expansion coefficients are the elements of the unitary matrix, ${\bf{U}}$, that diagonalizes ${\bf{H}}^\text{TC}$ ({\it{i.e.}}, $U_{im} = \beta_i^m$ if $i\le N$ and $U_{im} = \alpha_{i-N}^m$ if $i>N$) and hence orthogonal: $\sum_j^N\beta_j^{l*}\beta_j^m+\sum_k^{n_\text{modes}}\alpha_k^{l*}\alpha_k^m=\delta_{lm}$. Because the adiabatic states form a complete orthogonal set, the expansion of the total wave function in these adiabatic states is equivalent to the expansion in diabatic states (Equation~\ref{eq:Psi_dia}):
\begin{equation}
\Psi(t)=\sum_m^{N+n_\text{modes}}|\psi^m\rangle c_m(t)=\sum_j^{N+n_\text{modes}}|\phi_j\rangle d_j(t)\label{eq:Psiadia}
\end{equation}
with $c_m(t)$ the time-dependent adiabatic expansion coefficients. However, rather than propagating these adiabatic coefficients, as we did previously,\cite{Groenhof2019,Tichauer2021} we propagate in the diabatic basis instead, using the diabatic propagator (Equation~\ref{eq:propdia}). Because the diabatic and adiabatic states are connected by the unitary matrix {\bf{U}} (Equation~\ref{eq:adiabaticbasis}) via
\begin{equation}
d_j(t) = \sum_m^{N+n_\text{modes}} U_{jm}c_m(t)
\end{equation}
we can directly transform between the expansion coefficients of $|\Psi(t)\rangle$ in the two representations:\cite{Granucci2001}
\begin{equation}
    {\bf{c}}(t) = {\bf{U}}^{-1}{\bf{d}}(t)={\bf{U}}^{\dagger}{\bf{d}}(t)
\end{equation}
and obtain the propagator in the adiabatic basis:
\begin{equation}
\begin{array}{ccl}
{\bf{c}}(t+\Delta t) &=& {\bf{U}}^{\dagger}(t+\Delta t) {\bf{P}}^\text{dia}{\bf{U}}(t){\bf{c}}(t)\\
\\
&=& {\bf{P}}^\text{adia}{\bf{c}}(t)\label{eq:prop_adia}
\end{array}
\end{equation}
with ${\bf{P}}^\text{dia}$ the propagator in the diabatic basis, defined in Equation~\ref{eq:propdia}, and ${\bf{c}}$(t) the vector of adiabatic expansion coefficients ($c_m(t)$).

Thus, in this hybrid representation,\cite{Granucci2001} trajectories are propagated on a \emph{single} adiabatic potential energy surface $E_m$({\bf{R}}), associated with a polaritonic eigenstate $|\psi^m\rangle$ of $\hat{H}^\text{TC}$, until a hop takes place to the adiabatic potential energy surface $E_{i\neq m}$({\bf{R}}) of another polaritonic eigenstate, $|\psi^{i\neq m}\rangle$. Population is transferred from adiabatic state $|\psi^m\rangle$ into states $|\psi^{i\neq m}\rangle$ if
\begin{equation}
 | P^\text{adia}_{im}c_m(t)|^2>0
\end{equation}
where $P_{im}^\text{adia}=\langle\psi^i|\hat{P}^\text{adia}|\psi^m\rangle$ is the matrix element of the propagator in the adiabatic basis (Equation~\ref{eq:prop_adia}). To determine the probabilities for hopping from $|\psi^m\rangle$ to $|\psi^{i\neq m}\rangle$, we normalize $|P^\text{adia}_{im}c_m(t)|^2$ and multiply by the total probability to leave $|\psi^m\rangle$:\cite{Granucci2001}
\begin{equation}
  p_{m\rightarrow i} =%\frac{\delta |c_m|^2
  %}{|c_m(t)|^2}= 
  -\frac{ |c_m(t+\Delta t)|^2 - |c_m(t)|^2}{|c_m(t)|^2}
  \times \frac{|P_{im}^\text{adia}c_m(t)|^2}{\sum_{j\neq m}^{N+n_\text{modes}}|P_{jm}^\text{adia}c_m(t)|^2}\label{eq:FSSH}
\end{equation}
Next, we compare these probabilities to a random number. $\zeta$, drawn from a uniform distribution between 0 and 1, and decide to hop from adiabatic state $|\psi^m\rangle$ to another adiabatic state $|\psi^i\rangle$, if
\begin{equation}
  \sum_{q\neq m}^i p_{m\rightarrow
    q}<\zeta<\sum_{q\neq m}^{i+1}   p_{m\rightarrow q}\label{eq:random}
\end{equation}
After the hop, the evolution of the classical trajectory continues on the adiabatic potential energy surface, $E_i({\bf{R}})$, of state $|\psi^i\rangle$. 

To propagate the classical trajectory, we calculate the Hellmann-Feynmann forces on the atoms as expectation values of $\nabla\hat{H}^\text{TC}$ with respect to a single adiabatic eigenstate $|\psi^m\rangle$ of $\hat{H}^\text{TC}$ (Equation~\ref{eq:HTC}):\cite{Luk2017}
\begin{widetext}
\begin{equation}
  \begin{array}{ccl}
    {\bf{F}}^m_{a\in j}&=&-\langle \psi^m|\nabla_{a\in j}\hat{H}^{\text{TC}}|\psi^m\rangle \\
    \\&=&
    -\sum_j^N|\beta^m_j|^2\nabla_{a\in j}     \hat{H}^{\text{TC}}_{j,j}-\sum_k^{n_{\text{modes}}}|\alpha^m_k|^2\nabla_{a\in j} H^{\text{TC}}_{N+k,N+k}+\sum_j^N\sum_k^{n_{\text{modes}}}\beta_j^{m*}\alpha_k^m\nabla_a H^\text{TC}_{j,k+N}+\sum_k^{n_{\text{modes}}}\sum_j^N\alpha_k^{m*}\beta_j^m\nabla_{a\in j} H^{\text{TC}}_{k+N,j}\\
    \\
    &=& -|\beta_j^m|^2\nabla_{a\in j}V_{\text{S}_1}^\text{mol}({\bf{R}}_j)-\left(1-|\beta_j^m|^2\right)
    \nabla_{a\in
      j}V_{\text{S}_0}^\text{mol}({\bf{R}}_j)+2\Re\left[\beta_j^{m*}\nabla_{a\in j}{\boldsymbol{\mu}}^\text{TDM}_j({\bf{R}}_j)\cdot \sum_k^{n_{\text{modes}}}\alpha_k^m{\bf{f}}_k({\bf{R}}_j)  \right]\end{array}\label{eq:HellmannFeynman2}
\end{equation}
\end{widetext}
where we used the completeness of the adiabatic basis: $\sum_k^{n_\text{modes}}|\alpha_k^m|^2 = 1 -\sum_j^N\|\beta_j^m|^2$. As in Ehrenfest MD, the total wave functions is evolved along the classical trajectory by propagating the diabatic coefficient ${\bf{d}}(t)$ with the non-Hermitian propagator in the diabatic representation (${\bf{P}}^\text{dia}$, Equation~\ref{eq:propdia}). 

As is customary in Surface Hopping simulations, the total energy after a hop is conserved  by applying an {\it{ad hoc}} adjustment of the velocities.\cite{Hammes-Schiffer1994,Fang1999} Because the non-adiabatic coupling vector acts as a force that dissipates the energy gap between the adiabatic surfaces,\cite{Pechukas1969a,Pechukas1969b,Coker1995} velocities are adjusted along that vector.\cite{Tully91} The elements of the $3\times N_\text{mol}\times N_\text{atoms}$ dimensional non-adiabatic coupling vector ${\bf{D}}^{ml}$ for a hop from polaritonic eigenstate $|\psi_m\rangle$ to eigenstate $|\psi_l\rangle$ is computed as:\cite{Yarkony2012}
\begin{equation}
{\bf{D}}^{ml}_{a\in j} = \langle\psi^m | \nabla_{a\in j}|\psi^l\rangle = \frac{\langle \psi^m|\nabla_{a\in j}\hat{H}^{\text{TC}}|\psi^l\rangle}{E_l-E_m}\label{eq:nac}
\end{equation}
with $\nabla_{a\in j}$ the gradient with respect to the displacement of an atom $a$ in molecule $j$, and $E_l$ the adiabatic energy of polaritonic state $|\psi^l\rangle$. After substitution of the expression for the polaritonic eigenstates $|\psi^m\rangle$ (Equation~\ref{eq:Psiadia}), the Hellman-Feynman term in the numerator on the right-hand-side of Equation~\ref{eq:nac} becomes:\cite{Tichauer2022}
\begin{widetext}
    \begin{equation}
  \begin{array}{ccl}
    \langle \psi^m|\nabla_{a\in j}\hat{H}^{\text{TC}}|\psi^l\rangle
    &=&
    \beta_j^{m*}\beta_j^l\left[\nabla_{a\in j}V_{\text{S}_1}^\text{mol}({\bf{R}}_j)-\nabla_{a\in j}V_{\text{S}_0}^\text{mol}({\bf{R}}_j)\right]-\beta_j^{m*}\nabla_{a\in j}{\boldsymbol{\mu}}^\text{TDM}_j({\bf{R}}_j)\cdot
    {\bf{u}}_{\text{cav}}\sum_k^{n_{\text{max}}}\alpha_k^l
    {\bf{f}}_k({\bf{R}}_j)-\\
    \\
    &&
    \beta_j^l\nabla_{a\in j}{\boldsymbol{\mu}}^\text{TDM}_j({\bf{R}}_j)\cdot
    {\bf{u}}_{\text{cav}}\sum_k^{n_{\text{max}}}(\alpha_k^{m*}

    {\bf{f}}^*_k({\bf{R}}_j)
  \end{array}\label{eq:HellmannFeynman}
\end{equation}
\end{widetext}
Hops can only occur if the kinetic energy associated with the component of the total $3\times N_\text{mol}\times N_\text{atoms}$ dimensional momentum vector parallel to the non-adiabatic coupling vector, exceeds the energy gap between the adiabatic states. If there is sufficient kinetic energy, the velocity components parallel to the non-adiabatic coupling vector are adjusted as
\begin{equation}
   \dot{\bf{R}}_{a\in j}^\text{new} = \dot{\bf{R}}_{a\in j}^\text{old}-\xi
_{ml}{\bf{D}}^{ml}_{a\in j}/M_{a\in j}
\end{equation}
where $\dot{\bf{R}}_{a\in j}^\text{old}$ and $\dot{\bf{R}}_{a\in j}^\text{new}$  are the velocities of atom $a$ with mass $M_a$ in molecule $j$, before and after the hop from eigenstate $|\psi^l\rangle$ to $|\psi^m\rangle$, respectively. The factor $\xi_{ml}$ is obtained as the solution with the smallest absolute value of the quadratic equation:\cite{Hammes-Schiffer1994,Fang1999}
\begin{equation}
    \sum_i\frac{1}{2}M_i\left(\dot{\bf{R}}_{i}-\xi_{ml}\frac{{\bf{D}}^{ml}_{i}}{M_i}\right)^2=\sum_i\frac{1}{2}M_i\dot{\bf{R}}_{i}^2+E_m({\bf{R}})-E_l({\bf{R}})
\end{equation} 
If there is insufficient kinetic energy, the hop is aborted, but the components of the nuclear velocities parallel to the non-adiabatic coupling vector are reversed.\cite{Crespo2018}

\subsection{Multi-State Mapping Approach to Surface Hopping}

Because hops are stochastic, FSSH trajectories are not deterministic, which not only violates the principles of classical dynamics, but can also lead to inconsistencies between the potential energy surface on which the trajectory evolves and the polaritonic wavefunction. While such inconsistencies are normally overcome with decoherence corrections,\cite{Zhu2004, Granucci2007, Vindel-Zandbergen2021} these corrections are rather \textit{ad hoc} and sometimes lack a physical basis. To go beyond \textit{ad hoc} corrections, Mannouch and Richardson have proposed a novel approach to surface hopping, in which the classical trajectory always evolves on the potential energy surface of the adiabatic state with the highest population.\cite{Mannouch2023} Although originally derived for two coupled states, this Mapping Approach to Surface Hopping (MASH) was recently extended to multiple states by  Runeson and Manolopoulos.\cite{Runeson2023} Here, we implemented multi-state MASH for strongly coupled exciton-polariton systems. Instead of selecting the adiabatic potential energy surface by comparing computed hopping probabilities (Equation~\ref{eq:FSSH}) to a random number from a uniform distribution (Equation~\ref{eq:random}), we always select the surface $E_m({\bf{R}})$ of the adiabatic state $|\psi^m\rangle$ with the highest population in the total wavefunction, \textit{i.e.}, $m =\arg\max_{m} |c_m|^2$. As in FSSH, we adjust velocities after a hop to conserve energy and reverse velocities if there is insufficient kinetic energy for the hop.

\subsection{Explicit versus Implicit loss scheme}

% summary of the "implicity" decay model

Instead of accounting for the finite lifetime of cavity modes via the effective non-Hermitian Hamiltonian (Equation~\ref{eq:nonhermham}), which we refer to as an \emph{explicit} treatment of cavity losses, we had previously proposed an alternative approach, in which irreversible decay due to photon leakage through imperfect cavity mirrors was modelled by first-order decay of population in states with cavity mode contribution. Thus, the first-order rate at which adiabatic state $|\psi^m\rangle$ decays was computed as:\cite{Herrera2017PRL,Herrera2018}
\begin{equation}
\gamma_m = \sum_m^{n_\text{modes}}\gamma_k|\alpha_k^m|^2\label{eq:lossadia}
\end{equation}

While that approach, which we refer to as an \emph{implicit} treatment of cavity losses, was originally formulated for adiabatic eigenstates of the Tavis Cummings Hamiltonian (Equation~\ref{eq:HTC}),\cite{Tichauer2021} adaptation to a diabatic representation is straightforward. Indeed, because only diabatic states $|\phi_{N+k}\rangle$ that have an excitation of cavity mode $k$, can decay at a rate $\gamma_k$, the loss of population from such states during a MD time step $\Delta t$ is
\begin{equation}
    |d_{N+k}(t+\Delta t)|^2 =   |d_{N+k}(t)|^2 e^{-\gamma_k\Delta t}\label{eq:implicit_losses}
\end{equation}
Using that $|d_{N+k}|^2 = (\Re[d_{N+k}])^2 + (\Im[d_{N+k} ])^2$, we compute the change in the real and imaginary parts of the (complex) expansion coefficients $d_{N+k}(t)$ due to emission as:
\begin{equation}
  \begin{array}{ccc}
    \Re[d_{N+k} (t+\Delta t)] &=& \Re\left[d_{N+k}(t)\right] e^{-\frac{1}{2}\gamma_k\Delta
                              t}\\
    \\
    \Im[d_{N+k} (t+\Delta t)] &=& \Im\left[d_{N+k}(t)\right] e^{-\frac{1}{2}\gamma_k\Delta
                              t}\\
  \end{array}\label{eq:implicit_losses2}
\end{equation}
In  our implementation, cavity losses can thus be modeled either \emph{explicitly} by including the decay terms directly into the Hamiltonian (Equation~\ref{eq:nonhermham}), or \emph{implicitly} via exponential decay of populations in diabatic states with cavity mode excitation (Equation~\ref{eq:implicit_losses}).

\section{Simulation details}\label{section:simdetails}

To test the new implementation, and compare between including losses explicitly via the non-Hermitian Hamiltonian (Equation~\ref{eq:nonhermham}) or implicitly via a first-order decay process (Equation~\ref{eq:implicit_losses}) as in previous work ,\cite{Groenhof2019,Tichauer2021,Sokolovskii2023} we performed semi-classical MD simulations of the following processes:
\begin{enumerate}
    \item 
    Polariton relaxation in a multi-mode cavity, with a cavity loss rate of $\gamma_\text{cav}=$~66.7~ps$^{-1}$ ($\tau_\text{cav}=$~15~fs), in line with metallic cavities used experimentally;\cite{Schwartz2013}
    \item 
    Polariton transport in a multi-mode cavity with a loss rate of $\gamma_\text{cav}=$~66.7~ps$^{-1}$ ;
    \item Energy transfer in a hypothetical nano-cavity with cavity loss rates of  $\gamma_\text{cav}=$~200~ps$^{-1}$ ($\tau_\text{cav}=$~5~fs), in line with the lower Q factors of plasmonic nanoresonators.\cite{Lee2021}
\end{enumerate}
Before presenting the setup of these  molecule-cavity systems, we first share the details of the molecular models and simulation parameters used in these simulations.

\subsection{Molecular dynamics model systems}

% rhodamine in water

\subsubsection{Rhodamine in water}

\begin{figure}[!htb]
\centering
\includegraphics[width=0.4\textwidth ]{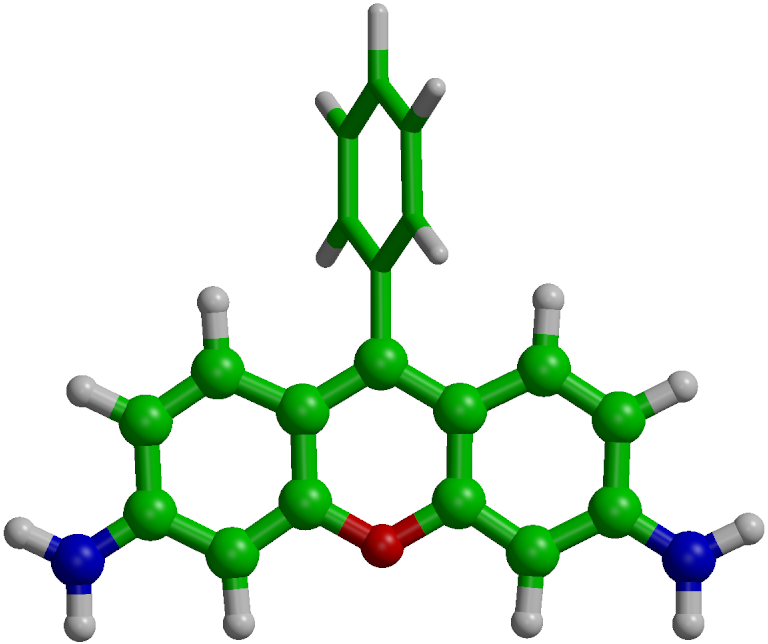}
  \caption{Rhodamine QM/MM model system. The QM atoms, described at the RHF/3-21G and CIS/3-21G levels of theory for the ground (S$_0$) and excited states (S$_1$), respectively, are shown in ball-and-stick representation, while the MM atoms, described with the Amber03 force field~\cite{Duan2003}, are shown as sticks. The hydrogen link atom introduced along the bond on the QM/MM interface to cap the valence of the QM subsystem is not shown and neither are the 3684 TIP3P water molecules.\cite{Jorgensen1983}
      }\label{fig:rhodamine}
\end{figure}

The Rhodamine molecule used in simulations 1 and 2, is shown in Figure~\ref{fig:rhodamine} and was modelled with the Amber03 force field,\cite{Duan2003} using the parameters derived by Luk {\it et al.}\cite{Luk2017} After a geometry optimization at the force field level, the molecule was placed at the center of a periodic rectangular box and filled with 3684 TIP3P water molecules.\cite{Jorgensen1983} The simulation box thus contained 11089 atoms and was equilibrated for 2~ns with harmonic restraints on the heavy atoms of the Rhodamine molecule (force constant 1000~kJmol$^{-1}$nm$^{-1}$). Subsequently, a 200~ns classical molecular dynamics (MD) trajectory was computed at constant temperature (300~K) using a stochastic dynamics integrator with a friction coefficient of 0.1~ps$^{-1}$. The pressure was kept constant at 1 bar using the Berendsen isotropic pressure coupling algorithm\cite{Berendsen1984} with a time constant of 1~ps. The LINCS algorithm was used to constrain bond lengths in the Rhodamine molecule,\cite{Hess1997} while SETTLE was applied to constrain the internal degrees of freedom of the water molecules,\cite{Miyamoto1992} enabling a time step of 2~fs in the classical MD simulations. A 1.0~nm cut-off was used for Van der Waals' interactions, which were modelled with Lennard-Jones potentials. Coulomb interactions were computed with the smooth particle mesh Ewald (PME) method,\cite{Essmann1995} using a 1.0~nm real space cut-off and a grid spacing of 0.12~nm. The relative PME tolerance at the real space cut-off was set to 10$^{-5}$. The simulations were performed with GROMACS~4.5.3.\cite{Hess2008}

The final configuration of the MM equilibration trajectory was subjected to a further 10~ps equilibration at the QM/MM level. The time step was reduced to 1~fs. As in previous work,\cite{Luk2017} the fused ring system was included in the QM region and described at the RHF/3-21G level of ab initio theory, while the rest of the molecule, as well as the water solvent were modelled with the Amber03 force field,\cite{Duan2003} and TIP3P water model,\cite{Jorgensen1983} respectively (Figure~\ref{fig:rhodamine}). The bond connecting the QM and MM subsystems was replaced by a constraint and the QM part was capped with a hydrogen atom. The force on the cap atom was distributed over the two atoms of the bond via the lever rule. The QM system experienced the Coulomb field of all MM atoms within a 1.6~nm cut-off sphere and Lennard-Jones interactions between MM and QM atoms were added. The singlet electronic excited state (S$_1$) of the QM region was modelled with the Configuration Interaction method, truncated at single electron excitations ({\it {i.e.,}} CIS/3-21G//Amber03). A comparison to more accurate (and costly) levels of theory in previous works~\cite{Groenhof2019, Sokolovskii2023} suggests that despite a significant overestimation of the excitation energy, CIS/3-21G yields potential energy surface topologies  that are in qualitative agreement with the more accurate approaches, including time-dependent density functional theory (TDDFT),\cite{Runge1984} complete active space self consistent field (CASSCF),\cite{Roos1999} and extended multi-configurational quasi-degenerate perturbation theory (xMCQDPT2).\cite{Granovsky2011} The QM/MM simulations were performed with GROMACS~4.5.3,\cite{Hess2008} interfaced to TeraChem.\cite{Ufimtsev2009,Titov2013} 

\subsubsection{Rhodamine and 10-hydroxybenzo[h]quinoline in DNA}

\begin{figure*}[!htb]
\centering
\includegraphics[width=\textwidth ]{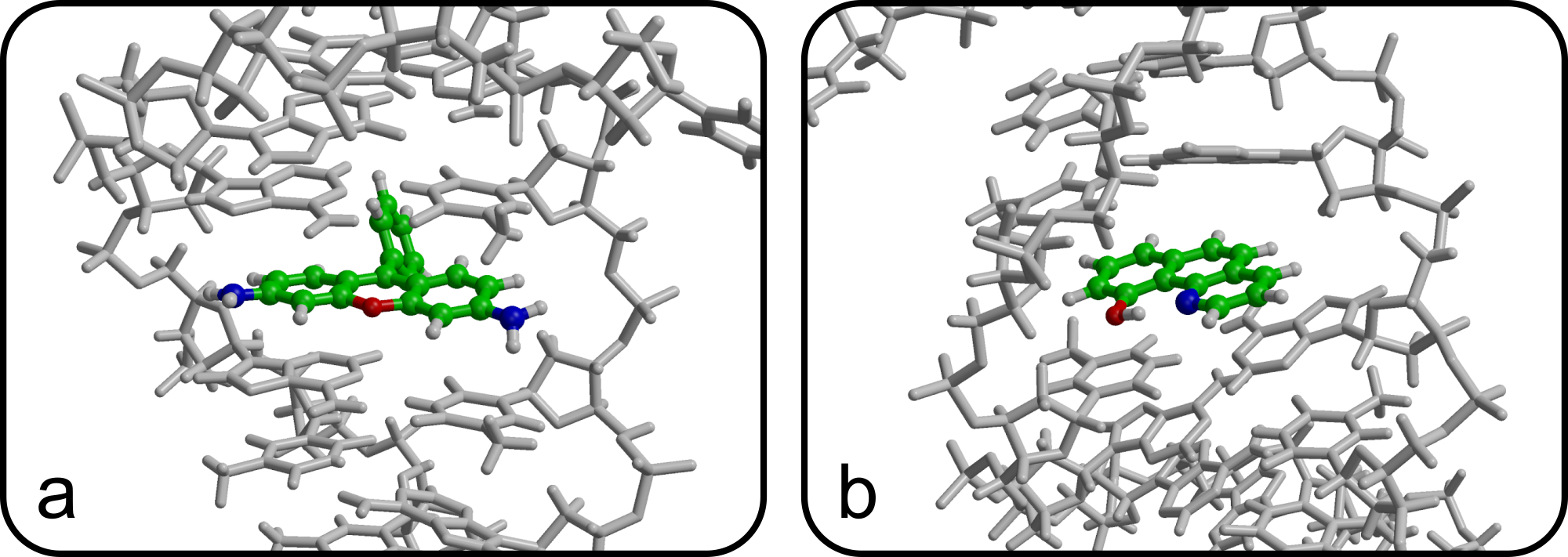}
  \caption{Rhodamine (\textbf{a}) and HBQ (\textbf{b}) intercalated in a DNA double helix (PDB ID: 1G3X). For Rhodamine the QM subsystem, shown in ball-and-stick in {\bf{a}}, was modelled at the HF/3-21G level of theory in the electronic ground state (S$_0$) and at the CIS/3-21G level of theory in the first singlet electronic excited state (S$_1$). For HBQ the QM subsystem, shown in ball-and-stick in {\bf{b}}, was modelled at the CAM-B3LYP/6-31G(d) DFT level in the electronic ground state (S$_0$) and at the TDA-CAM-B3LYP/6-31G(d) time-dependent DFT level in the first singlet electronic excited state (S$_1$). The MM subsystem, consisting of the DNA (shown in gray in stick representation), as well as the water solvent with Na$^+$ cations and Cl$^-$ anions (not shown), was modelled with the Amber99-SB force field.}
\label{fig:Rho-HBQ}
\end{figure*}

Because double-stranded DNA can be used to self-assemble a nano-plasmonic cavity,\cite{Heintz2021} we built a hypothetical model of such cavity, in which the DNA not only maintains the structural integrity of the nano-cavity, but also contains two different chromophores intercalated between base pairs. This cavity model was used for the simulations of polariton-assisted energy transfer between the intercalated chromophores. The initial structure for the DNA in these simulations is the x-ray structure of the DNA / N$^\alpha$-(9-acridinoyl)-tetra-arginine intercalation complex (PDB ID: 1G3X).\cite{Malinina2002} The acridine-peptide drug was replaced by a Rhodamine molecule in one structure and by 10-hydroxybenzo[h]quinoline (HBQ) in another, via a least-squares fit of the dyes onto the drug (Figure~\ref{fig:Rho-HBQ}). The interactions between the atoms in these systems were modelled with the Amber99-SB force field. \cite{Hornak2006} For Rhodamine, we used the same Amber atom types as before,\cite{Luk2017} while for HBQ, we used atom type {\tt{CA}} for the aromatic carbons, {\tt{HC}} for the aromatic hydrogens, {\tt{NC}} for the nitrogen, {\tt{OH}} for the hydroxyl oxygen and {\tt{HO}} for the hydroxyl hydrogen.

% figure of Rhodamine and HBQ in DNA

After molecular replacement, the DNA-chromophore complexes were energy minimized using the limited-memory Broyden-Fletcher-Goldfarb-Shanno algorithm (l-BFGS).\cite{Liu1989} The energy-minimized DNA-Rhodamine complex was placed in a  rectangular periodic box and solvated with 8427~TIP3P water molecules.\cite{Jorgensen1983} To keep the system neutral at 0.2~M ion concentration, 37~Na$^+$ and 16~Cl$^-$ ions were added. The energy-minimized DNA-HBQ complex was placed in a rectangular  periodic box as well, to which 12176 water molecules, 45 Na$^+$ and 23 Cl$^-$ ions were added. 

The boxes were equilibrated for 10~ns at constant temperature (300~K) and pressure (1 bar) using the v-rescale thermostat ($\tau_T$=0.1 ps$^{-1}$),\cite{Bussi2007} and the Berendsen isotropic pressure coupling algorithm ($\tau_p =$~1~ps$^{-1}$),\cite{Berendsen1984} respectively. The LINCS algorithm was used to constrain bonds involving hydrogen atoms,\cite{Hess1997} while SETTLE was applied to constrain the internal degrees of freedom of the water molecules,\cite{Miyamoto1992} enabling a time step of 2~fs in the classical MD simulations. During equilibration the coordinates of the Rhodamine and HBQ atoms were kept fixed. Van der Waals interactions were modelled with Lennard-Jones potentials, truncated at 1.0~nm, while electrostatic interactions were modelled with the smooth PME method,\cite{Essmann1995} using a 1.0~nm real-space cut-off and a grid spacing of 0.12~nm. The relative tolerance at the real-space cut-off was set to 10$^{-5}$. The simulations were performed with GROMACS~4.5.3.\cite{Hess2008}

The equilibration trajectories were continued for 10~ps at the QM/MM level with a time step of 1~fs. In these simulations, the DNA, water molecules and ions were modelled with the Amber99-SB force field,\cite{Hornak2006} while the complete Rhodamine molecule was modelled at the RHF/3-21G level of theory in electronic ground state (S$_0$) and at the CIS/3-21G level in the electronic excited state (S$_1$). The electronic ground (S$_0$) and excited (S$_1$) states of the HBQ molecule were modeled with Density Functional Theory (DFT),\cite{Hohenberg1964} and time-dependent density functional theory (TDDFT),\cite{Runge1984} within the Tamm-Dancoff approximation (TDA),\cite{Hirata1999} respectively, using the CAM-B3LYP functional,\cite{Becke1993,Yanai2004} in combination with the 6-31G(d) basis set.\cite{Ditchfield1971} The QM subsystems experienced the Coulomb field of all MM atoms within a 1.6 nm cut-off sphere and Lennard-Jones interactions between MM and QM atoms were added. The QM/MM simulations were performed with GROMACS~4.5.3,\cite{Hess2008} interfaced to TeraChem.\cite{Ufimtsev2009,Titov2013} 

\subsection{Molecular dynamics of cavity-molecule systems}

\subsubsection{Polariton relaxation} 
After QM/MM equilibration, we placed 64 Rhodamine molecules, including solvent, with equal intermolecular spacings on the $z$-axis of a periodic one-dimensional (1D),\cite{Michetti2005,Tichauer2021} 5~$\mu$m long, optical Fabry-P\'erot micro-cavity. The dispersion of this cavity, $\omega_\text{cav}(k_{z,p})=\sqrt{\omega^2_0+ c^2k_{z,p}^2}$, was modelled  with 16~discrete modes ({\it i.e.}, $k_{z,p}=2\pi p/L_z$ with $0\leq p\leq 16$ and $L_z=$~5 $\mu$m). The micro-cavity was red-detuned with respect to the Rhodamine absorption maximum, which is 4.18~eV at the CIS/3-21G//Amber03 level of theory, such that the energy of the fundamental mode at normal incidence ($k_z=0$) was $\hslash\omega_0=3.81$ eV, corresponding to a distance of $L_x=$~0.163 $\mu$m between the mirrors. With a cavity vacuum field strength of 0.0002 au (1.0~MVcm$^{-1}$), the Rabi splitting was  $\sim$325~meV. We assumed a radiative decay rate of $\gamma_\text{cav} = $ 66.7~ps$^{-1}$ for all cavity modes. Because for the Rhodamine model employed in this work, the S$_1$/S$_0$ conical intersection is about 1 eV higher in energy than the Franck-Condon region on the S$_1$ potential energy surface,\cite{Sokolovskii2023} and the typical nano-second lifetime of molecular excitations is several orders of magnitude longer than that of the cavity modes (15~fs), we neglected radiationless deactivation of the molecules and assumed an infinite excited-state lifetime for the molecules instead.

After instantaneous excitation into the 68$^\text{th}$ eigenstate, which corresponds to a point on the UP branch, Ehrenfest MD trajectories\cite{Ehrenfest1927} were computed by numerically integrating Newton's equations of motion using a leap-frog algorithm with a 0.1~fs time step.\cite{Verlet1967} We performed three simulations, in which (\textit{i}) we propagate the polaritonic wave function in the diabatic basis with loss terms added explicitly to the effective non-hermitian Hamiltonian (Equation~\ref{eq:nonhermham}); (\textit{ii}) we propagate the polaritonic wave function with the Hermitian Hamitonian in the diabatic basis with losses implicitly modelled as first-order decay of the populations (Equation~\ref{eq:implicit_losses}); and (\textit{iii}) we propagate the polaritonic wave function in the adiabatic basis with losses implicitly modelled as first-order decay of the populations.\cite{Tichauer2021} 

To facilitate the comparison between the different propagation schemes, the three simulations were started with identical initial atomic coordinates and velocities. The initial coordinates were the same for all Rhodamines, whereas the initial velocities were selected randomly from a Maxwell-Boltzmann distribution at a temperature of 300~K. To maximize the coupling strength, the molecules were oriented such that their transition dipole moments aligned with the polarization of the vacuum field inside the cavity at the start of the simulations. The simulations were performed with GROMACS~4.5.3,~\cite{Hess2008} in which the multi-mode Tavis-Cummings QM/MM model was implemented,\cite{Tichauer2021} in combination with TeraChem.\cite{Ufimtsev2009,Titov2013} 

\subsubsection{Polariton transport}

We placed 1024~Rhodamine molecules, including water, at equal intermolecular intervals on the $z$-axis of a 1D,\cite{Michetti2005} 50~$\mu$m long, optical Fabry-P\'erot micro-cavity. As in previous work on polariton propagation,\cite{Agranovich2007,Sokolovskii2023,Tichauer2023} the dispersion of this cavity was modelled  with 160~discrete modes ({\it i.e.}, $k_{z,p}=2\pi p/L_z$ with $0\leq p\leq 159$ and $L_z=$~50 $\mu$m). The fundamental mode was red-detuned by 370~meV with respect to the excitation energy of Rhodamine (4.18~eV at the CIS/3-2G//Amber03 level of theory). Thus, the energy at normal incidence was  $\hslash\omega_0=3.81$ eV,  corresponding to a distance of $L_x=$ 0.163~$\mu$m between the mirrors. With a cavity vacuum field strength of 0.00005 au (0.26~MVcm$^{-1}$), the Rabi splitting was $\sim$325~meV. The same decay rate of $\gamma_\text{cav} = $ 66.7~ps$^{-1}$ was used for all cavity modes, whereas an infinite lifetime was assumed for the molecules.

In experiments, polariton propagation is often initiated via off-resonant excitation into a higher-energy electronic state of a single molecule.\cite{Lerario2017,Berghuis2022,Balasubrahmaniyam2022} Under the assumption that the subsequent relaxation into the S$_1$ state of that molecule is ultra-fast,\cite{Kasha1950} we modeled such off-resonant excitation conditions by starting the simulations with one of the molecules, $j$, in the S$_1$ electronic state: $\sigma^+_j|\phi_0\rangle$. We computed Ehrenfest MD trajectories using a leap-frog algorithm with a 0.1~fs time step. \cite{Verlet1967} To test the new possibilities of our implementation, we performed four sets of simulations: (\textit{i}) in the diabatic representation with explicit cavity losses; (\textit{ii}) in the diabatic representation with implicit losses; (\textit{iii}) in the adiabatic representation with implicit losses and (\textit{iv}) in the hybrid diabatic/adiabatic representation with explicit losses. 

The four simulations were started with identical initial atomic coordinates and velocities. The initial coordinates were the same for all Rhodamines, whereas the initial velocities were selected randomly from a Maxwell-Boltzmann distribution at a temperature of 300~K. At the start of the simulations, the molecules were oriented to maximize the coupling strength by aligning their transition dipole moments to the polarization of the vacuum field inside the cavity. The simulations were performed with GROMACS~4.5.3,\cite{Hess2008} in which the multi-mode Tavis-Cummings QM/MM model was implemented,\cite{Tichauer2021} in combination with Gaussian16.\cite{g16}

\subsubsection{Polaritonic energy transfer}

A DNA double strand containing an intercalated Rhodamine molecule, with an excitation energy at 4.04~eV, and a DNA double strand containing an intercalated HBQ molecule with an excitation energy at 4.16~eV, were coupled together to the same single-mode cavity tuned at 4.11~eV, with a cavity vacuum field strength of 0.001 au (5~MVcm$^{-1}$) and decay rate of $\gamma_\text{cav}=$ 200~ps$^{-1}$ ($\tau_\text{cav}=$ 5~fs). The hypothetical nano-plasmonic cavity is thus modelled implicitly. In future work, we will aim at including the metal nanoparticles explicitly into the MM region, using a suitable metal nano-particle force field.\cite{Pohjolainen2016}. %Two sets of simulations were performed: one with a cavity decay rate of $\gamma_\text{cav}=$ 66.7~ps$^{-1}$ ($\tau_\text{cav}=$ 15~fs) and the other with a faster decay rate of  $\gamma_\text{cav}=$ 200~ps$^{-1}$ ($\tau_\text{cav}=$ 5~fs). 
We performed FSSH simulations with and without the decoherence correction of Granucci \textit{et al.},\cite{Granucci2007} as well as MASH simulations without decoherence correction. For the decoherence correction in the FSSH simulations, we used the default parameter of 0.1 Hartree. For each of these simulations, two series of 100 trajectories were computed, one in which the lower polariton (LP) is initially excited and another set in which the upper polariton (UP) is initially excited. The polaritonic wave function (Equation~\ref{eq:adiabaticbasis}) was propagated in the diabatic basis, while the molecular dynamics on the adiabatic surfaces were integrated with a 0.5~fs time step. The starting coordinates and velocities were sampled at 100~fs intervals from the ground state QM/MM trajectories. All simulations were run for 100~fs with Gromacs~4.5.3,\cite{Hess2008} in which the Tavis-Cummings QM/MM model was implemented, in combination with TeraChem.\cite{Ufimtsev2009,Titov2013} 

\section{Results and Discussion}\label{section:results}

\subsection{Polariton relaxation}

\begin{figure*}[!htb]
\centering
\includegraphics[width=\textwidth]{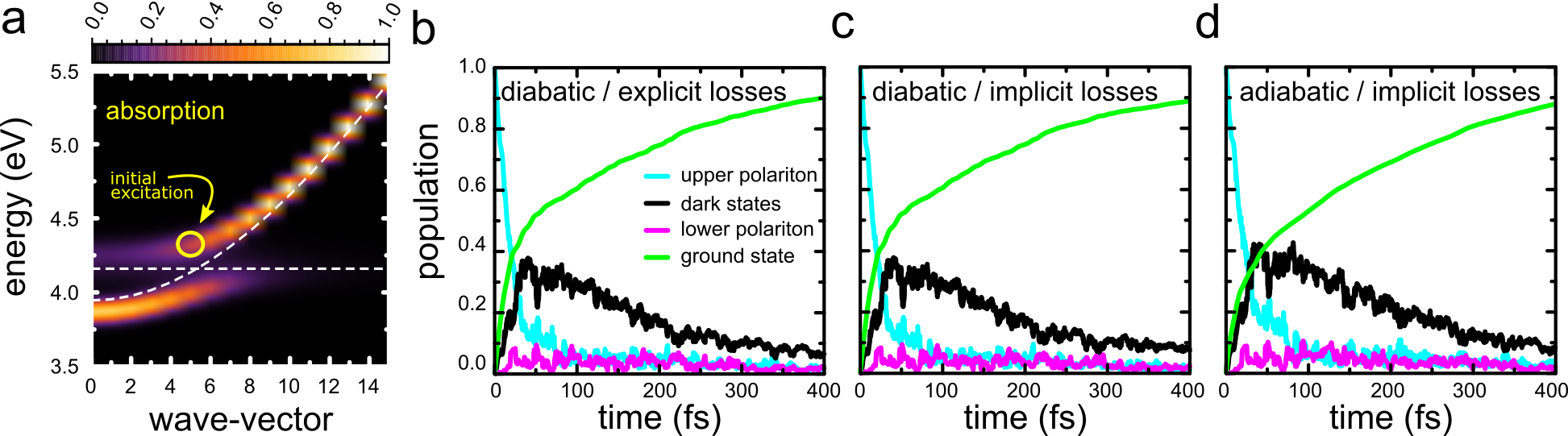}
  \caption{Absorption spectrum of 64~Rhodamine molecules, including water environment, in a 1D Fabry-P\'{e}rot micro-cavity modelled with 16~discrete modes ({\bf{a}}). Panels {\bf{b}}-{\bf{d}}: Population dynamics after an instantaneous resonant excitation into an eigenstate in the upper polariton branch, indicated by the yellow circle in panel~{\bf{a}}. The polaritonic wave function was coherently propagated in the diabatic representation with explicit losses~({\bf{b}}), in the diabatic representation with implicit losses~({\bf{c}}), and in the adiabatic representation with implicit losses~({\bf{d}}), while the nuclear degrees of freedom were evolved on the mean-field potential energy surface.}
\label{fig:relax}
\end{figure*}

In Figure~\ref{fig:relax}, we show the population dynamics in a lossy cavity containing 64~Rhodamines after an instantaneous resonant excitation with a hypothetical narrow-band delta pulse into an eigenstate of the UP branch, indicated by the yellow circle in the angle-resolved absorption (or visibility\cite{Lidzey1999}) spectrum of panel {\bf a}. In these simulations, the total polaritonic wave function (Equation~\ref{eq:Psiadia})
was propagated in the diabatic representation with cavity decay treated explicitly (Figure~\ref{fig:relax}{\bf{b}}) or implicitly (Figure~\ref{fig:relax}{\bf{c}}). For comparison, we also plot the populations when the wave function is propagated in the adiabatic representation using the implicit treatment of the losses (Figure~\ref{fig:relax}{\bf{d}}), as in Tichauer {\it et al.}~\cite{Tichauer2021}

In line with results from previous quantum mechanical and semi-classical simulations,\cite{DelPino2018,Groenhof2019,Tichauer2021} and consistent with earlier theoretical findings,\cite{Agranovich2003,Litinskaya2004,Mazza2009} population is rapidly transferred from the initially excited UP state into the dark states, which we define as eigenstates of the Tavis-Cummings Hamiltonian, $|\psi^m\rangle$, for which the total contribution of the 16~cavity mode excitations (Equation~\ref{eq:Psiadia}) is below a threshold, {\it{i.e.}}, $\sum_{k=1}^{n_{\text{modes}}}|\alpha_n^m|^2 < 0.05$. The dynamics of the populations, including that of the ground state (green line in Figure~\ref{fig:relax}), is very similar for the two simulations in the diabatic representation, suggesting no major differences in treating cavity losses explicitly with the effective non-Hermitian Hamiltonian (Figure~\ref{fig:relax}{\bf{b}}),\cite{Ulusoy2019,Felicetti2020,Antoniou2020} or implicitly as a first-order decay process of populations in diabatic states with cavity mode contributions ({\it{i.e.}}, $|\phi_j\rangle$, with $j>N$, Equation~\ref{eq:implicit_losses}, Figure~\ref{fig:relax}{\bf{c}}). 

In contrast, comparing the simulations in the diabatic basis with an explicit or implicit treatment of the cavity losses on the one hand, to the simulation in the adiabatic representation with implicit losses (Figure~\ref{fig:relax}{\bf{d}}) on the other hand, reveals small differences in the population dynamics, with a faster rise of dark state population initially and a concomitant slower decay into the ground state for the simulation in the adiabatic basis with implicit losses. As we will discuss in more detail below, this difference is due to how we had modelled the implicit decay for multiple cavity modes within the adiabatic representation.\cite{Tichauer2021} 

\subsection{Polariton transport}

We performed four simulations of 1024~Rhodamine molecules strongly coupled to 160~confined light modes of an unidirectional 1D Fabry-P\'{e}rot cavity. As in previous work,\cite{Sokolovskii2023} we modeled the off-resonant excitation of the molecule-cavity system, in which a higher-energy electronic excited state localized on a single molecule is pumped, which then rapidly relaxes into the lowest-energy electronic excited state,\cite{Kasha1950} by starting the simulations with one molecule (located at 5 $\mu$m) in the S$_1$ electronic excited state. In these simulations, the total polaritonic wave function (Equation~\ref{eq:Psiadia}) was coherently propagated in (\textit{i}) the diabatic basis with implicit losses; (\textit{ii} and \textit{iii}) the diabatic basis with explicit losses; and (\textit{iv}) the adiabatic basis with implicit losses. The classical MD trajectory was evolved on the mean-field potential energy surface in the diabatic representation for simulations \textit{i} and \textit{ii}, and in the adiabatic representation for simulations \textit{iii} and \textit{iv}. 

\begin{figure*}[!htb]
\centering
\includegraphics[width=0.96\textwidth ]{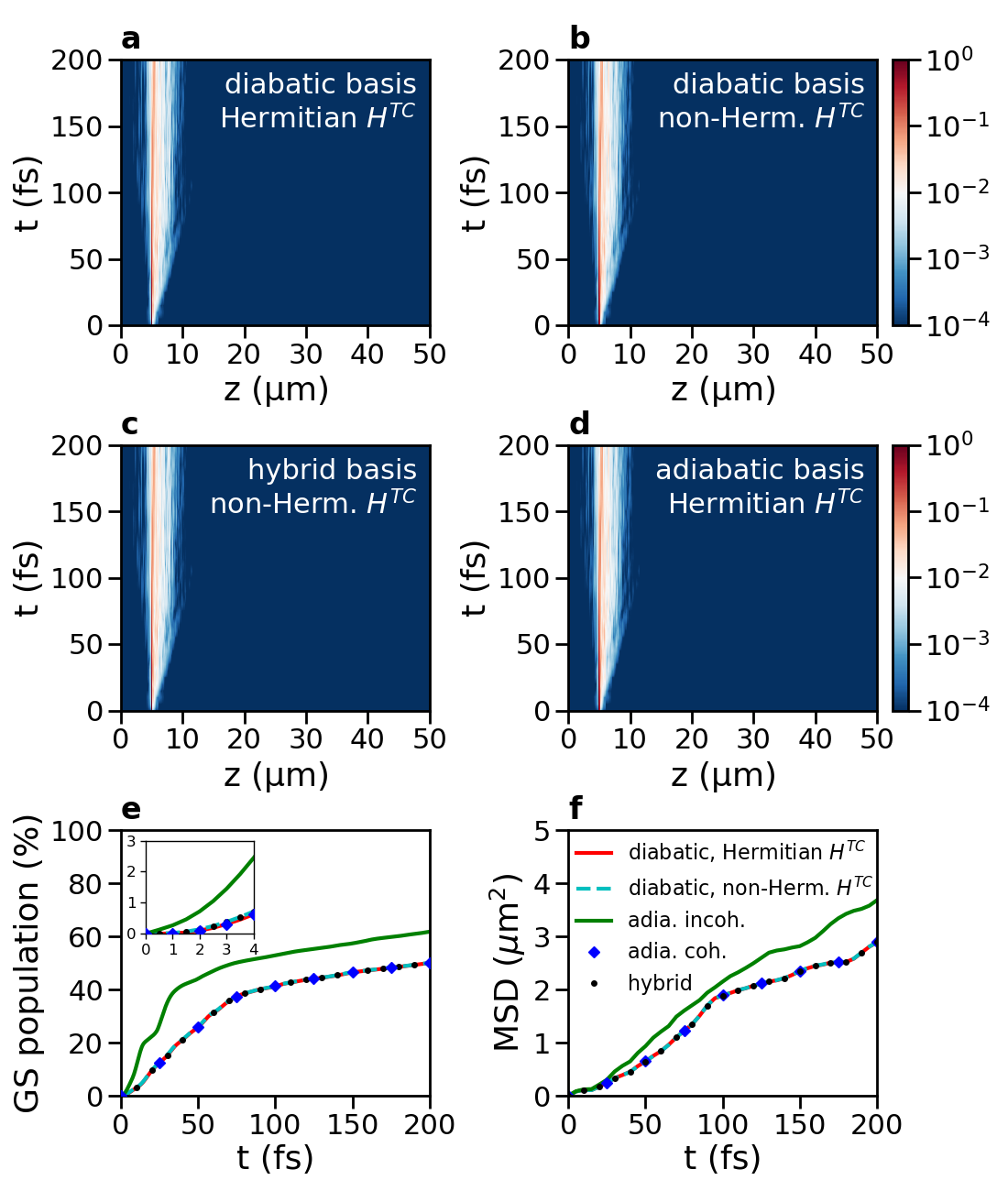}
  \caption{Polariton propagation after  instantaneous excitation of molecule $j$ at position $z_j=5$~$\mu$m, into the S$_1$ electronic state (\textit{i.e.}, $\sigma^+_j|\phi_0\rangle$). Panels (\textbf{a})--(\textbf{d}) depict space-time maps of the probability density $\vert\Psi(z,t)\vert^2$ of the polariton wave function, in a lossy cavity ($\gamma_{\text{cav}}=$~66.7~ps$^{-1}$) containing 1024~Rhodamine chromophores. The polaritonic wave function was evolved in the diabatic representation with both implicit (\textbf{a}) and explicit (\textbf{b}) inclusion of losses, the hybrid representation (\textit{i.e.},  propagating the wave function in the diabatic basis, while propagating the trajectory in the adiabatic basis) with explicit cavity decay (\textbf{c}), and the adiabatic representation with the implicit loss scheme (\textbf{d}). Panels (\textbf{e}) and (\textbf{f}) show the total ground state (GS) population and the mean squared displacement (MSD) of the total polariton wave function in the simulations.
  %calculated with the same propagation schemes. %as well as when losses as summed incoherently in the adiabatic representation. % during the whole simulation time, as well as first 4~fs (inset). 
  The inset in panel (\textbf{e}) shows the evulaiton of ground state population in the first 4~fs.}
\label{fig:transport}
\end{figure*}

In Figure~\ref{fig:transport}\textbf{a}--\textbf{d} we show the time evolution of the probability density of the total polaritonic wave function, $|\Psi(z,t)|^2$ (Equation~\ref{eq:Psiadia}). In all simulations, we observe that after the instantaneous off-resonant excitation of a single molecule, a propagating wavepacket forms due to population transfer from the S$_1$ state of that molecule into bright polaritonic states with group velocity. Initially, that population transfer is mostly driven by Rabi oscillations, as the initial state is not an eigenstate of the strongly coupled molecule-cavity system, but on longer timescales population transfer continues due to thermally driven molecular displacements of vibrational modes that overlap with the non-adiabatic coupling vector.\cite{Tichauer2022,Sokolovskii2023} Because these population transfers are reversible,\cite{Groenhof2019} and hence also occur from the propagating bright states back into the stationary dark state manifold, the propagation appears as a diffusion process,\cite{Sokolovskii2023} in line with experimental observations.\cite{Rozenman2018}

Because the polariton transport mechanism was investigated and discussed in detail in previous works,\cite{Berghuis2022,Sokolovskii2023,Tichauer2023} we here focus on the differences between the four propagation schemes. In simulations \textit{ii} and \textit{iii}, in which the wave function was propagated in the diabatic basis with the effective non-Hermitian Hamiltonian, the wavepacket propagation is the same (panels {\bf b} and {\bf{c}} in Figure~\ref{fig:transport}), irrespective of what basis is used to evaluate the mean-field forces for the evolution of the classical trajectory. Consequently, the decay into the ground state (\textit{i.e.}, $\rho_0(t)=1-\sum_m|d_m(t)|^2$) is identical (Figure~\ref{fig:transport}{\bf{e}}) in these two simulations and also the Mean-Squared-Displacement (MSD) of the total wavepacket, $|\Psi(z,t)|^2$ is the same (Figure~\ref{fig:transport}{\bf{f}}). 

With an implicit treatment of radiative losses as first-order decay of population from diabatic states representing cavity mode excitations ({\it i.e.}, $|\phi_{j>N}\rangle$, Equation~\ref{eq:implicit_losses}), in simulation \textit{i} (Figure~\ref{fig:transport}{\bf a}) the wavepacket propagates as in simulations in which the losses were included explicitly into an effective non-Hermitian Hamiltonian (\textit{ii} and \textit{iii}). The total loss, or ground state population at the end of the simulation, only deviates on the order of 0.1~\%  (solid red line in Figure~\ref{fig:transport}\textbf{e}), while the MSDs are indistinguishable (Figure~\ref{fig:transport}\textbf{f}). These observations suggest that including losses implicitly, while keeping the Hamiltonian Hermitian, provides a viable alternative for the propagation with a non-Hermitian Hamiltonian, in particular for larger systems, for which the inversion of a non-hermitian matrix can be computationally more demanding than the diagonalization of an Hermitian matrix.

However, if we propagate the wavefunction in the basis of adiabatic eigenstates (simulation \textit{iv}), and treat the losses as in previous work by incoherently summing the cavity mode contributions to a polaritonic state to obtain the total decay rate of that state (\textit{i.e.}, $\gamma_m^\text{tot} = \sum_k^{n_\text{modes}}\gamma_k|a_k^m|^2$),\cite{Herrera2017PRL,Tichauer2021} we observe differences in the evolution of the wave packet as compared to simulations \textit{i}, \textit{ii} and \textit{iii} (Figure~\ref{fig:transport}{\bf{d}}). As shown in panels {\bf e} and {\bf f} of Figure~\ref{fig:transport}, the radiative decay is faster initially, and also the MSD increases more steeply, suggesting a higher diffusion coefficient. Closer inspection of the evolution of the ground state population (inset in Figure~\ref{fig:transport}{\bf e}) reveals that the decay rate at the start of simulation \textit{iii} is non-zero. In contrast, for the simulations in which the wave function was propagated in the diabatic representation,  the decay rate is zero initially, which is consistent with starting in a state that is fully localized onto a single molecule and therefore has no cavity mode contributions. 

Because the initial state is the same in both representations (\textit{i.e.}, $|\Psi(0)\rangle^{\text{loc. on~} j}=\sum_m c_m|\psi^m\rangle=\sum_m\beta_j^m|\psi^m\rangle=|\phi_j\rangle$), the difference is due to how the loss rates are computed. Whereas the loss rate at the start of the simulation is zero in the diabatic representation, as the initial coefficients for diabatic states with cavity mode excitation are zero ({\it i.e.}, $\dot{\rho}_0(t)=\sum_k^{n_\text{modes}}\gamma|d_{N+k}(t)|^2=0$), summing 
incoherently over cavity mode contributions to compute the decay rate of the eigenstates $|\psi^m\rangle$ in the adiabatic expansion of $|\Psi(0)\rangle$ (Equation~\ref{eq:lossadia}),\cite{Herrera2017PRL,Tichauer2021} can lead to initial decay rates that are higher than zero: $\dot{\rho}_0(t) = \gamma\sum_m\sum_k|\alpha_k^m|^2|c_m(t)|^2=\gamma\sum_m|\beta_j^m|^2\left(\sum_k|\alpha_k^m|^2\right)\ge 0$. To confirm that these differences are due to the incoherent summing of the cavity mode contributions to the adiabatic states, we repeated the simulation in the adiabatic basis, but transform these states into diabatic states before computing the losses implicitly with Equation~\ref{eq:implicit_losses}. Indeed, when losses are modeled this way (blue diamonds in Figure~\ref{fig:transport}{\bf{e}},{\bf{f}}), propagation in the adiabatic basis becomes fully consistent with propagation in the diabatic basis. Furthermore, also when simulations are initiated in bright polaritonic states, either by direct excitation into a bright eigenstate ( Figure~\ref{fig:relax}), or by resonant excitation into a linear combination of bright polaritonic states with a broad band laser pulse (Figure~S1 in Supporting Information), propagation in adiabatic basis with implicit losses, calculated as before,\cite{Tichauer2021} yields similar decay as propagation in the diabatic basis with either implicit or explicitly losses. 

%of the wave function with explicit losses yield very similar decay rates. %Thus, treating the losses explicity in our previous simulations of polariton transport under on-resonant excitation conditions, 
%Nevertheless, in our new implementation, the losses can be handled consistently, irrespective of the representation used to provide the potential energy surface for the classical molecular dynamics.

Summarizing, the results of our simulations suggest that treating radiative losses into the far-field explicitly by adding decay terms to cavity mode energies in the Hamiltonian, or implicitly via an exponential decay of population in diabatic states that represent the cavity modes, has no major impact when the propagation is done in the diabatic representation. However, the comparisons revealed that the approach we had proposed previously for treating losses in adiabatic eigenstates,\cite{Tichauer2021} can overestimate the losses if the initial state is not an eigenstate of the molecule-cavity Hamiltonian.

\subsection{Energy transfer}

To test the implementation of fewest-switches surface hopping with explicit losses for systems with more than one molecule, we simulated a Rhodamine and a HBQ molecule collectively coupled to a single-mode nano-cavity. Here, we assume that the cavity is formed as in Heintz \textit{et al.} via self-assembly of two complementary DNA strands, both of which are covalently attached to a metal nano-particle.\cite{Heintz2021} While in that experimental work, also the seven Atto-647N dye molecules with which the strong-coupling regime was reached, were linked covalently to the DNA, the Rhodamine and HBQ in our simulations are assumed to enter the cavity mode volume by forming non-covalent intercalation complexes with the DNA through $\pi$-stacking with the base pairs (Figure~\ref{fig:Rho-HBQ}).\cite{Malinina2002} In these simulations, the highly dissipative metal nanoparticles that form the actual nano-cavity in Heintz {\it{et al.}}\cite{Heintz2021} were not included, but  modelled implicitly instead as a single lossy cavity mode with a vacuum field strength of 0.001~au (5.1~MVcm$^{-1}$) and lifetime of 5~fs. These simulations are a first step towards a more sophisticated and fully atomistic model of dyes in such plasmonic nano-cavities in future work. 

\begin{figure}[htb]
\centering
\includegraphics[width=0.4\textwidth ]{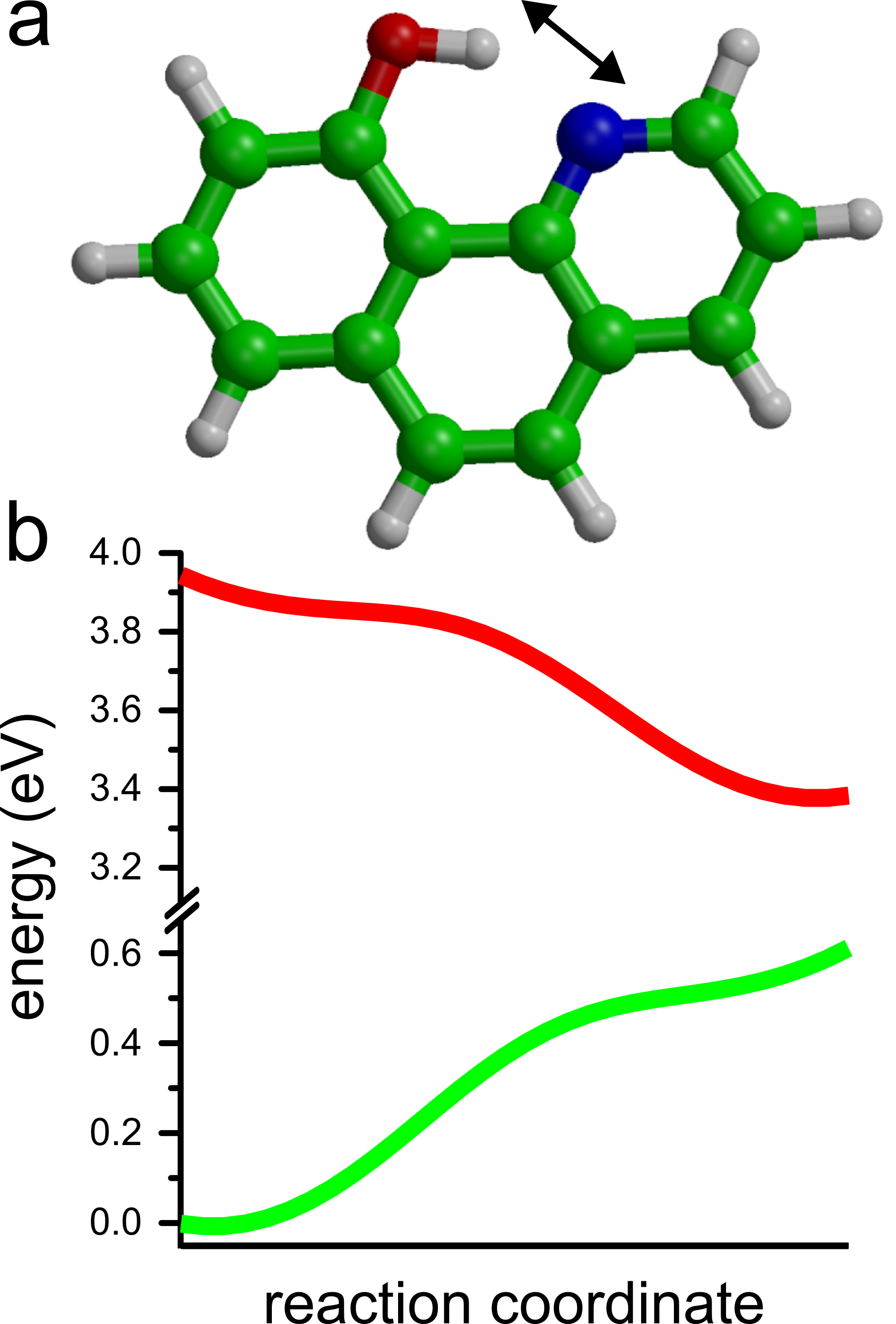}
  \caption{HBQ (\textbf{a}) and potential energy profiles for proton transfer in the electronic ground (S$_0$, green) and excited state (S$_1$, red), evaluated at the CAMB3LYP/6-31G(d) level of theory (\textbf{b}). The reaction coordinate, indicated by the double arrow in panel \textbf{a}, is defined as the difference between Oxygen-Hydrogen and Nitrogen-Hydrogen distances: $d_\text{O-H} - d_\text{N-H}$.
  }\label{fig:HBQ}
\end{figure}

In addition to testing the FSSH implementation, the key question we want to address here is whether exciting into one of the optically-accessible polaritonic states, which are separated by a Rabi splitting of $\hbar\Omega^\text{Rabi}=$ 236.1~meV (Figure~S2, in Supporting Information), can lead to efficient energy capture by HBQ despite the high loss rate of the cavity mode. As photo-excited HBQ undergoes an ultra-fast proton transfer reaction into a photo-product that cannot couple to the cavity due to a red-shift of over 1~eV (Figure~\ref{fig:HBQ}),\cite{Lee2013} this energy capture process can be conveniently tracked by monitoring the distance between the hydroxyl oxygen atom and the proton.

\begin{figure*}[!htb]
\centering
\includegraphics[width=\textwidth]{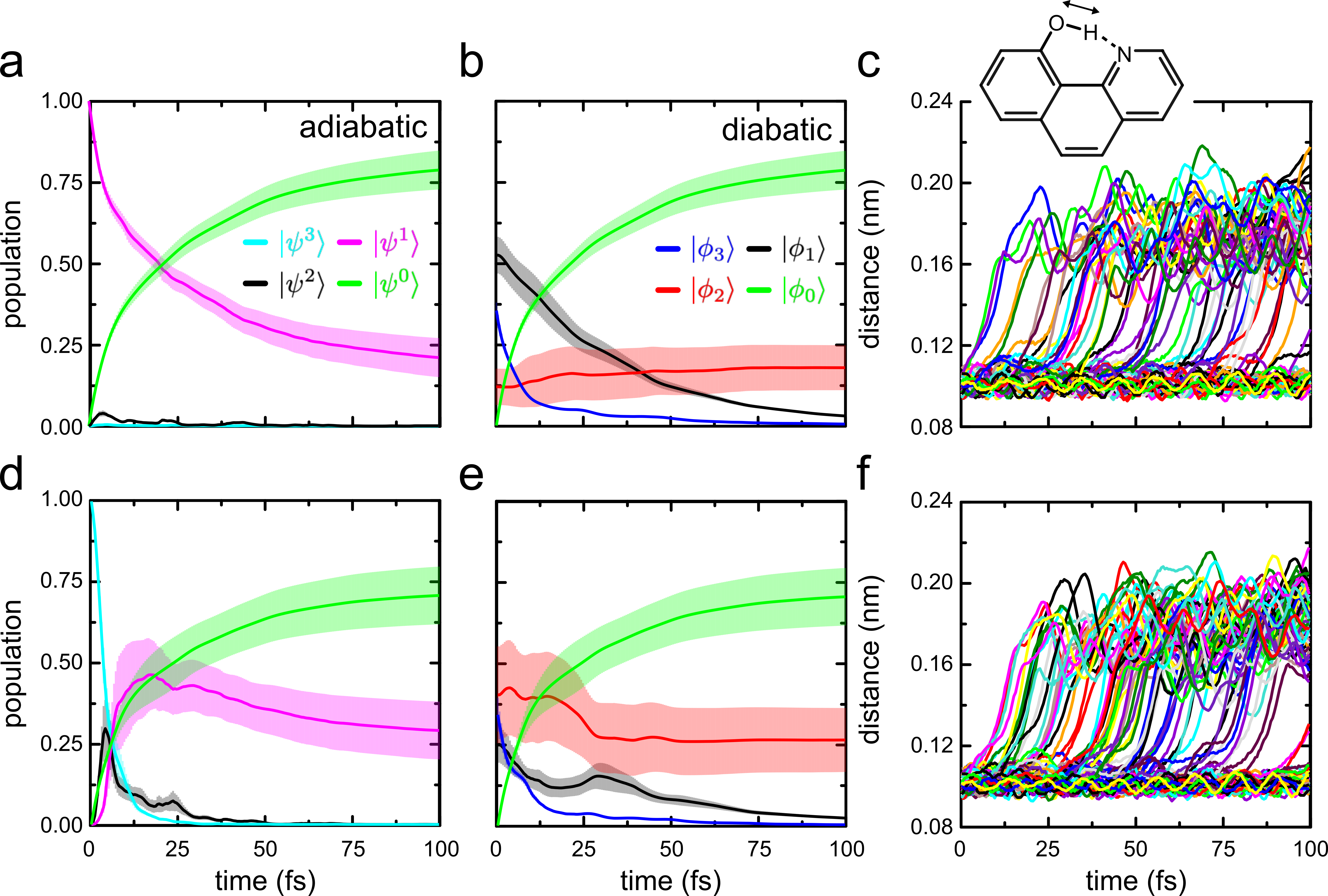}
  \caption{Top panels: Average population of the adiabatic ($|\psi^m\rangle$, panel {\bf{a}}) and diabatic states ($|\phi_j\rangle$, panel {\bf b}
) and distance between the hydroxyl oxygen and proton (double arrow in the inset, panel {\bf c}) after instantaneous excitation into the adiabatic lower polariton state. The diabatic states are defined as: $|\phi_1\rangle=|\text{S}_1^\text{Rho}\text{S}_0^\text{HBQ}\rangle\otimes|0\rangle$; $|\phi_2\rangle=|\text{S}_0^\text{Rho}\text{S}_1^\text{HBQ}\rangle\otimes|0\rangle$; and $|\phi_3\rangle=|\text{S}_0^\text{Rho}\text{S}_0^\text{HBQ}\rangle\otimes|1\rangle$. Bottom panels: Average population of the adiabatic ($|\psi^m\rangle$, panel {\bf d}) and diabatic states ($|\phi_j\rangle$, panel {\bf e}) and distance between the hydroxyl oxygen and proton (panel {\bf f}) after instantaneous excitation into the adiabatic upper polariton state. The ground state population ($|\psi^0\rangle=|\phi_0\rangle=|\text{S}_0^\text{Rho}\text{S}_0^\text{HBQ}\rangle\otimes|0\rangle$) is shown in green. Distances in panels {\bf c} and {\bf d} are shown for each of the 100~simulations separately with different colours, whereas the populations in panels {\bf a}, {\bf b}, {\bf d} and {\bf e}  were averaged over the 100~simulations. The transparent areas around the curves in the average population plots show the root-mean-squared deviation ({\it i.e.}, $\sigma_\rho = \sqrt{\langle \rho_m^2\rangle-\langle \rho_m\rangle^2}$ ).}
\label{fig:energy_transfer}
\end{figure*}

In panel {\bf{a}} of Figure~\ref{fig:energy_transfer}, we show the  populations of the adiabatic states ($\rho_m(t) = |c_m(t)|^2$), including the total ground state ($\rho_0(t) = 1-\sum_m|c_m(t)|^2$), averaged over hundred simulations after excitation into the lowest-energy eigenstate of the system, which is a polariton with a 52\%  excitonic contribution of Rhodamine ({\it i.e.}, $|\beta_1^1|^2 = $ 0.52 with $|\phi_1\rangle = |\text{S}_1^\text{Rho}\text{S}_0^\text{HBQ}\rangle\otimes|0\rangle$) and 35\% of the cavity photon ($|\alpha^1|^2 = 0.35$ with $|\phi_3\rangle = |\text{S}_0^\text{Rho}\text{S}_0^\text{HBQ}\rangle\otimes|1\rangle$). As shown in Figure~\ref{fig:energy_transfer}{\bf{c}}, HBQ undergoes intra-molecular proton transfer in 46 out of 100 simulations. Alternatively, exciting into the highest-energy eigenstate (lower panels in Figure~\ref{fig:energy_transfer}), which is a polariton with a $\sim$34\% contribution from the cavity photon, and excitonic contributions from both HBQ ($\sim$41\%) and Rhodamine ($\sim$25\%), leads to proton transfer in 63 simulations, suggesting that in this system, exciting into the UP provides a more efficient route for transforming the photon energy into chemical energy than the LP, in line with quantum dynamics simulations on NaI.\cite{Vendrell2018b} 

However, because the cavity mode is very lossy, radiative decay into the far field, indicated by the rise in ground state population (green line in Figure~\ref{fig:energy_transfer}), competes with the proton transfer reaction and reduces the quantum yield, defined as the average population in diabatic state $|\phi_2\rangle = |\text{S}_0^\text{Rho}\text{S}_1^\text{HBQ}\rangle\otimes|0\rangle$ at the end of the simulation, to 18 $\pm$ 7\% for excitation into LP and to 27 $\pm$ 10\% for excitation into the UP. Repeating these simulations with MASH instead of FSSH yields highly similar results (Figure~S4 in Supporting Information). While, here, the purpose of the simulations was to test our implementation and verify that we can run FSSH simulations in the collective strong coupling regime, future work will be aimed at including an atomistic description of the metal nano-particles into the MM subsystem,\cite{Pohjolainen2016} as well as a more accurate description of the quantized electro-magnetic fields.\cite{Medina2021}

\section{Conclusion}\label{section:conclusion}
  
In summary, to model the effect of cavity decay in atomistic molecular dynamics simulations of collectively coupled exciton-polaritons, we have implemented the effective non-Hermitian Hamiltonian,\cite{Ulusoy2020,Antoniou2020,Felicetti2020,Hu2022} in which radiative losses into the far field are added as imaginary contributions to the cavity mode energy terms. To keep the potential energy surfaces on which the classical trajectories evolve real, we implemented the hybrid diabatic / adiabatic semi-classical molecular dynamics approach of Granucci and co-workers,\cite{Granucci2001} in which the polaritonic wave function is propagated in the diabatic basis under the influence of the effective non-hermitian Hamiltonian, while the classical nuclei move on an single adiabatic potential energy surface, or linear combinations thereof. We have shown that with the new implementation, we can simulate the dynamics of large ensembles of molecules collectively coupled to the lossy modes of an optical cavity, and investigated relaxation, transport and energy transfer. The addition of the diabatic representation and the effective non-Hermitian Hamiltonian to our multi-scale MD approach paves the way for more advanced descriptions of the cavity mode structure, such as few-mode quantisation, which requires an explicit inclusion of cavity losses into the Hamiltonian.\cite{Medina2021,Sanchez-Barquilla2022}

\begin{acknowledgments}
We thank Oriol Vendrel for pointing us to the advantages of performing our simulations in the diabatic basis and very valuable discussions. We furthermore thank Tero Heikkil\"{a}, Pauli Virtanen, Johannes Feist, Pavel Buslaev, Dmitry Morozov, Ruth H. Tichauer and Fedor Nigmatulin for sharing their insights into various aspects of this work. We also thank Ruth H. Tichauer for feedback on the code. The CSC-IT center for scientific computing in Espoo, Finland, is acknowledged for providing very generous computational resources. This work was supported by the Academy of Finland (Grants 323996 and 332743).
%\dots.
\end{acknowledgments}

\section*{Data Availability Statement}

The data that support the findings of this study are available from the corresponding author upon reasonable request. The code, based on a fork of Gromacs-4.5.3, is available for download at https://github.com/upper-polariton/GMXTC.git

\section*{Supporting Information}
Details on initial conditions in the transport simulations; wavepacket analysis; computation of absorption spectra; additional results for (\textit{i}) polariton transport under resonant excitation  conditions, (\textit{ii}) absorption spectrum of the Rhodamine/HBQ nanocavity system, and (\textit{iii}) MASH simulations of energy transfer in the Rhodamine/HBQ nanocavity system. 

\bibliography{main}
%\bibliography{aipsamp}% Produces the bibliography via BibTeX.

\end{document}